\documentclass[9pt,twocolumn]{extarticle}
\usepackage[margin=0.75in]{geometry}
\usepackage{microtype}

\usepackage{titlesec}
\titleformat{\section}
  {\normalfont\bfseries\large}{}
  {0pt}{}
\titleformat{\subsection}
  {\normalfont\bfseries\itshape\normalsize}{}
  {0pt}{}
\titlespacing*{\section}{0pt}{1.2ex}{0.6ex}
\titlespacing*{\subsection}{0pt}{1.0ex}{0.4ex}

\usepackage{caption,graphicx}
\usepackage{amssymb}
\usepackage{tablefootnote}
\usepackage{etoolbox}

\usepackage[autostyle, english=american]{csquotes}
\MakeOuterQuote{"}

\usepackage{enumitem}
\setlist[itemize]{
  leftmargin=*,
  labelsep=0.6em,
  label=\raisebox{0.28ex}{$\scriptstyle\blacktriangleright$},
  itemsep=0.1ex,
  parsep=0pt,
  topsep=0.2ex
}

\usepackage{xspace}
\newcommand{\eg}{\textit{e.g.},\xspace}

\usepackage[nolist]{acronym}
\begin{acronym}
  \acro{LCA}{life cycle assessment}
  \acro{LCT}{life cycle thinking}
  \acro{TRL}{technology readiness level}
  \acro{EoL}{end-of-life}
  \acro{AI}{artificial intelligence}
  \acro{LCI}{life cycle inventory}
  \acrodefplural{LCI}{life cycle inventories}
  \acro{LCIA}{life cycle impact assessment}
  \acro{i/o}{input/output}
  \acro{GHG}{greenhouse gas}
  \acro{ML}{machine learning}
  \acro{TEA}{techno-economic analysis}
  \acro{SbD}{sustainable-by-design}
\end{acronym}

\usepackage{booktabs}
\usepackage[table,xcdraw]{xcolor}
\usepackage[hidelinks]{hyperref}
\usepackage{cleveref,xr}
\usepackage{xurl}
\usepackage[english]{babel}
\newcommand{\boldref}[1]{%
  \ifcsdef{used@#1}{\autoref{#1}}{%
    \global\csdef{used@#1}{}%
    \textbf{\autoref{#1}}%
  }%
}

\usepackage[
  backend=biber,
  style=nature,
  sorting=none
]{biblatex}
\AtEveryBibitem{%
  \clearfield{note}%
  \clearfield{month}%
  \clearfield{url}%
  \clearfield{urldate}%
  \clearfield{urlmonth}%
  \clearfield{urlyear}%
}
\let\cite\supercite

\begin{document}

\title{\textbf{Sustainability-informed materials design}}


\author{
Rachel Woods-Robinson\textsuperscript{1},
Amalie Trewartha\textsuperscript{2}\\[0.3em]
{\small
\parbox{0.9\textwidth}{\centering
\textsuperscript{1}Clean Energy Institute, University of Washington, Seattle, WA 98195, United States\\
\textsuperscript{2}Toyota Research Institute, Los Altos, CA 94022, United States
}}
}

\date{\today}

\twocolumn[

\maketitle
\begin{center}
\begin{minipage}{\textwidth}
\begin{abstract}
While material innovation can enable sustainable development, environmental and social impacts of emerging materials are often assessed only after design choices are “locked in.” Here, we argue for a shift in perspective: life cycle thinking should enter at the earliest stages of materials development, where uncertainty is highest but design freedom is greatest. Rather than treating incomplete knowledge as a barrier, we reframe it as an inherent feature that can illuminate trajectories, tradeoffs, and consequences—and enable intervention while change remains possible. Focusing on inorganic solid materials, we identify disconnects between materials science and sustainability analysis, propose an adaptable, decision-oriented framework to embed sustainability into material design across evolving technology stages, and highlight how recent advances such as predictive synthesis can help operationalize this integration. Guided by the framework’s governing principles, we outline a cross-stakeholder agenda to shift from retrospective correction to anticipatory, responsible material design from the outset.\\ 
\end{abstract}
\end{minipage}
\end{center}
]

\raggedbottom
\addtolength{\textheight}{-1\baselineskip} 

\section{Motivation}

The transition to a low-carbon future will require technologies enabled by advances in functional materials, yet the sustainability of these new materials is not guaranteed from the outset. Early design choices can have cascading consequences, shaping downstream impacts that are difficult to mitigate once systems scale. An illustrative example is lithium-ion battery cathode materials, where cobalt-use has led to social and environmental harms, and delayed industry response has necessitated costly redesigns.\cite{li_cobalt_2020, GlobalCriticalMinerals2025} Because early choices in the research cycle constrain future sustainability outcomes, responsible deployment of new materials will require designing for reduced adverse impacts alongside performance and cost.

\begin{figure}[htb!]
    \centering
    \includegraphics[width=0.5\textwidth]{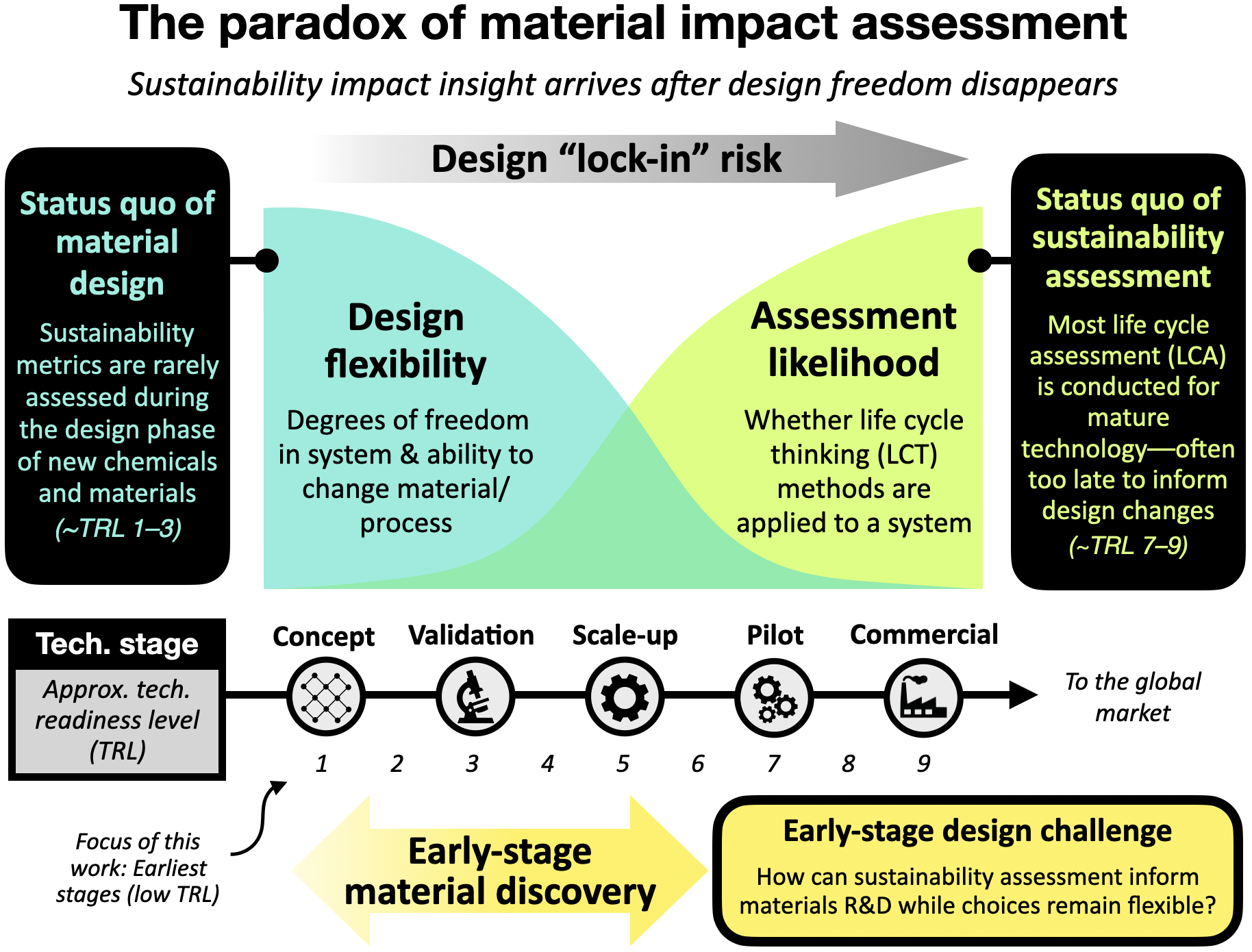}
    \caption{Schematic representation of the current state of sustainability impact assessment in inorganic solid-state materials across various \acfp{TRL}. "Design flexibility" reflects the degrees of freedom in material and process choices, while "assessment likelihood" reflects the practical feasibility and commonness of practice of applying \acf{LCA} or related \acf{LCT} methods. \ac{TRL} positions and curve shapes are illustrative and not intended to represent precise, quantitative relationships.}
    \label{fig:paradox}
\end{figure}

\Ac{LCT} and its quantitative tool, \ac{LCA}, provide a comprehensive framework for evaluating environmental burdens from extraction through end-of-life.\cite{iso_iso_2006, iso_iso_2006-1} However, as illustrated in \boldref{fig:paradox}, \ac{LCA} is typically applied retrospectively at from the pilot stage onward, where harmful impacts may be "locked in" and redesign is costly.\cite{arvidsson_environmental_2018} \ac{LCA} is often perceived as impractical at low \acp{TRL} due to high uncertainty, limited process data, and unknown scale-up trajectories. Yet at low-\acp{TRL} this uncertainty reflects substantial remaining design freedom---precisely where assessment can still inform scale-up decisions and meaningfully influence outcomes. For emerging technologies, uncertainty and decision relevance are thus inseparable.

In parallel, recent years have seen the emergence of data-driven workflows for inorganic solid-state materials discovery, increasingly coupled with \ac{ML} and autonomous experimentation.\cite{otyepka_advancing_2025, tom_self-driving_2024, tobias_autonomous_2025} These developments are moving the field toward \textit{inverse design}, marked by searches over vast design spaces for both candidate materials with targeted properties and for plausible pathways to synthesize them. In principle, predictive synthesis tools could generate candidate reaction pathways, precursors, and processing conditions to help address data gaps and construct the \acp{LCI} used as input to early-stage \ac{LCA}; while retrosynthesis-informed LCA workflows are emerging in organic and green chemistry,\cite{mikolajczyk_retrosynthesis_2023, blanco_machine_2024, chen_life_2026} analogous integration remains limited for inorganic solids.

Here, we argue there is both an opportunity and a need to break down disciplinary silos between materials discovery and sustainability assessment. We propose a framework and concrete actions to integrate \ac{LCT} into materials design from the earliest stages, (\textasciitilde\ac{TRL} 1–3), where choices about composition, synthesis, and performance targets remain most flexible---and thus most impactful. Early focus also enables a reframing of sustainability assessment itself: rather than relying on conventional \ac{LCA} constraints of well-defined data, we advocate for an \ac{LCT} approach that starts from just the basic information available at discovery and leverages uncertainty to define tunable design spaces. As technologies mature, this approach can be iteratively refined with increasing data and fidelity, ultimately converging towards the conventional \ac{LCA} common at higher TRLs. Our discussion highlights environmental impacts of solid-state materials, but its principles are broadly applicable across materials, impact categories, and complementary methods such as \ac{TEA}.

\section{Vision: Life Cycle Thinking as a Design Principle}

We propose a framework to embed \ac{LCT} in materials discovery at the point of material selection to inform concrete research decisions, rather than deferring until key design choices are fixed. \autoref{fig:framework} outlines how \ac{LCT} can guide early-stage materials design while remaining compatible with higher-fidelity assessment as technologies mature, and is constructed to align with the five core principles.

\begin{enumerate}
    
    \item \textbf{Bottom-up construction.} At early TRLs, sustainability assessment should begin from information available  at discovery (or inferrable from first-principles), \eg composition, structure, plausible synthesis pathways. In contrast to a conventional top-down \ac{LCA} approach---requiring complete process flows, supply chains, operating conditions, and use-phase details---a bottom-up approach builds outward from the material itself and treats unknowns as scenarios or bounds. This makes assessment actionable for discovery-stage research, where multiple pathways remain open and knowledge is necessarily partial but still informative.
    
    \item \textbf{Uncertainty as information.} Rather than a reason to defer analysis, uncertainty at low TRLs can be reframed as an \textit{explicit modeling feature}. Applied as a source of insight, uncertainty bounds and scenarios can reveal dominant drivers and leverage points, even when precise estimates are not possible. Framed this way, uncertainty becomes productive: it can help define tunable design spaces and guide where additional R\&D focus could most effectively reduce impacts.
    
    \item \textbf{Enabling progressive, iterative refinement across \acp{TRL}.} Early-stage \ac{LCT}, while not itself a full LCA, should be structured to preserve compatibility with high-\ac{TRL} assessments by aligning with ISO-defined methodological steps\cite{iso_iso_2006, iso_iso_2006-1}---goal and scope, \ac{LCI}, \ac{LCIA}, and interpretation---and conventional life cycle phases. This \textit{allows continuous evolution towards a full LCA} as technologies mature, with assumptions progressively refined, proxies replaced by measured data, and uncertainty narrowed through iteration rather than model replacement. As \ac{TRL} increases, the role of assessment shifts from broad screening and bounding to pathway-specific comparison and higher-fidelity evaluation, reframing sustainability assessment as an iterative loop that co-evolves with design.

    \item \textbf{Decision relevance over precision.} The purpose of low-\ac{TRL} \ac{LCT} is distinct from that of high-\ac{TRL} \ac{LCA}: it is not to produce or certify definitive impact values, but to \textit{inform choices} and support ranking and prioritization among materials, pathways, and research directions. Emphasizing decision relevance over numerical precision allows sustainability considerations to shape which options are explored and which tradeoffs are pursued, without imposing inappropriate expectations of accuracy under deep uncertainty.
    
    \item \textbf{Integration through interoperable, modular design.} Materials databases, process models, and life-cycle inventories are fragmented across platforms, use incompatible taxonomies, and are often subject to access restrictions. Interoperability, shared ontologies, and open access are therefore essential for materials informatics and \ac{LCT} tools to work together, and to enable a shift from static datasets toward \textit{dynamic, science-grounded models} that can evolve as knowledge improves.

\end{enumerate}

Together, these principles define an approach to \ac{LCT} that treats sustainability as a performance metric---alongside functionality and cost---to inform early design decisions without constraining exploration.

\section{Enabling Advances: From Principles to Practice}

\begin{figure*}[htb!]
    \centering
    \includegraphics[width=1\textwidth]{fig2-framework.png}
    \caption{A vision and framework for \acf{LCT} in materials design. (a) Comparison of conventional material R\&D's focus on performance and cost (i) with a proposed pathway integrating \ac{LCT} into early-stage materials design (ii). Conventional workflows occasionally consider sustainability \textit{ad hoc} checks (\eg "earth-abundance" or toxicity, based on elemental composition), but rarely consider process-level implications across the life cycle. The proposed pathway introduces a dedicated life-cycle–informed design step. (b) Modular framework for incorporating \ac{LCT} into early-stage materials design, with tasks organized according to the four ISO-standardized phases of \ac{LCA}\cite{iso_iso_2006, iso_iso_2006-1}: goal and scope (system boundaries, functional units, and available process parameters), \acf{LCI} (candidate synthesis routes; estimated material and energy flows), \acf{LCIA} (impact quantification; note most \ac{LCA} focuses on environmental impact factors, but emerging social \ac{LCA} is shown here for reference) and interpretation (contextualizing results). Tasks within each phase need not be applied simultaneously; early-stage analyses may focus on a subset sufficient to bound dominant drivers. The schematic highlights representative considerations for low \acp{TRL} and is not exhaustive---tasks may be added, refined, or omitted as technologies mature and new data becomes available (\eg microstructure unlikely relevant at \ac{TRL} 1; reaction prediction unnecessary once synthesis is demonstrated).}
    \label{fig:framework}
\end{figure*}

Recent developments mark a shift towards the feasibility of early-stage \ac{LCT}, enabled not by a single breakthrough but by a convergence of advances across disciplines. \boldref{fig:framework} outlines a practical implementation of our \ac{LCT} framework, intended as modular rather than prescriptive (\eg only a subset of tasks might be considered at low TRLs). It begins with a modified ISO-defined "goal and scope" phase (\autoref{fig:framework}b, phase 1) to define the decision context and establish bounds on material parameters (\eg available precursors). Here, we highlight recent advances that enable implementation of the framework---especially in \ac{LCI} setup, decision-relevant assessment, and data integration---and identify remaining challenges.

\nopagebreak[4]

\subsection{Early-Stage Life Cycle Inventory Construction}

\nopagebreak[4]

The primary challenge of estimating impacts for new materials without fully specified process data is translation: mapping structure-level materials information to plausible material and energy flows without a fully specified process. Here, we introduce how a foreground \acp{LCI} could be bottom-up constructed by assembling bounded, scenario-based models of material and energy flows from what is knowable at discovery (see \autoref{fig:framework} panel 2). A key enabling factor is the rapid acceleration of materials informatics and generative modeling.

While synthesis routes were historically established through experimental trial-and-error, new techniques now allow plausible processing pathways to be enumerated much earlier.\cite{szczypinski_can_2021, kovnir_predictive_2021, aziz_towards_2021, leeman_challenges_2024} Starting from just the target material (\eg a crystal structure), as well as constraints such as available precursors and synthesis methods, predictive synthesis can propose potential reaction sequences and thermodynamic processing ranges (\eg temperature, pressure). Approaches in inorganic solids span from physics-based models (\eg reaction networks)\cite{aykol_network_2019, aykol_rational_2021, mcdermott_graph-based_2021, gallant_reactca_2024} to ML-based methods (\eg graph, sequence, diffusion, large language models),\cite{malik_predicting_2021, huo_machine-learning_2022, szymanski_autonomous_2023, szymanski_autonomous_2023-1, williamson_using_2023, kim_predicting_2024, kim_large_2024, chen_data-driven_2026, pan_diffsyn_2026} and can support explicit uncertainty quantification.(\eg ensemble or Bayesian approaches)\cite{mcdermott_graph-based_2021, gallant_reactca_2024, kim_predicting_2024} These models---as well as literature-, patent-, and data-mined synthesis routes for known materials\cite{kononova_text-mined_2019, kim_inorganic_2020, makino_extracting_2021, wang_dataset_2022, chung_solid-state_2025}---are increasingly published as open datasets, providing access without needing to run models, but work is needed to standardize synthesis terminology and representations.\cite{wang_ulsa_2022} Crucially, these reactions could be used to define \ac{i/o} flows sufficient to construct \acp{LCI}, yet they remain underutilized in \ac{LCT} workflows.

Beyond stoichiometric reactions, the energetic transformations required to achieve a specific material form (\eg purity, morphology, or microstructure) often drive life cycle impacts. For example, in silicon solar cells, purification to solar-grade (>99.999999\% pure Si\cite{semi_semi_2019}) and melting into crystalline ingots dominate energy and carbon intensity.\cite{frischknecht_life_2020} Conventional process simulation tools (\eg Aspen Plus) can model energy in chemical processes for \ac{LCI},\cite{parvatker_simulation-based_2020} but often rely on proprietary environments and rigid process definitions that limit interoperability with both materials and \ac{LCA} platforms.\cite{lee_idaes_2021} Integrating estimates into end-to-end workflows would require open, scalable tools that leverage reaction energetics, phase behavior, and process heuristics---such as surrogate,\cite{alizadeh_managing_2020, misener_formulating_2023} \ac{ML},\cite{brumen_overview_2021, schweidtmann_machine_2021, schweidtmann_mining_2024} and other open process modeling tools (\eg DWSIM,\cite{medeiros_dwsim_2025} biosteam,\cite{cortes-pena_biosteam_2020} idaes\cite{lee_idaes_2021})---to provide physically-grounded bounds on processing energy that can be incorporated as \ac{i/o} flows.

As technologies advance through TRLs, experimental measurements can progressively replace or refine these modeled bounds. Alternatively, when research begins in the lab rather than via computation, documenting synthesis routes in inventory-relevant terms---such as precursor choices, reaction stoichiometry, yields, and processing conditions---could allow build-up of similar \ac{i/o} models, and evolution alongside process development and scale-up aligned with prospective \ac{LCA} methods. In this setting, inventory-informed synthesis optimization could also be explored, \eg through autonomous labs or Bayesian-inference approaches that iteratively refine synthesis conditions while evaluating impact implications.\cite{shields_bayesian_2021}

Once candidate \ac{i/o} models are defined, parametric foreground \acp{LCI} can explicitly encode uncertainty into stoichiometry, yields, energy-intensities, or process-level efficiencies. Recent developments across prospective \ac{LCA}\cite{arvidsson_environmental_2018, thonemann_how_2020, langkauStepwiseApproachScenariobased2023} and statistical proxy methods (\eg underspecification\cite{olivetti_exploring_2013}) allow \ac{LCI} construction as bounded ranges, analog processes, and scenario assumptions rather than as fully specified unit operations. Scenario-based and probabilistic representations can propagate uncertainty across alternative synthesis routes, processing conditions, and learning trajectories, preserving multiple plausible inventories (\eg scale-up scenarios).\cite{erakcaSystematicReviewScaleup2024a, nurdiawatiRecentAdvancementsProspective2025a}

\subsection{Decision-Relevant Life Cycle Impact Assessment (LCIA) and Interpretation}

Existing \ac{LCIA} methods can be applied to any \ac{LCI}, including those derived from early-stage, highly uncertain data. As in conventional \ac{LCA}, the challenge is typically not the impact calculation itself---since mapping the \ac{LCI} to impact indicators is largely embedded within standard \ac{LCA} software---but rather in selecting which impact categories (\eg \ac{GHG} emissions, water use) are most relevant and interpreting results under deep uncertainty.\cite{hauschild_identifying_2013}

A unique challenge of novel materials is that their environmental and health interactions are not well characterized, and therefore likely lack established characterization factors, the coefficients used in \ac{LCIA} to translate inventory flows into impact indicators (\eg climate or toxicity impacts).\cite{cucurachi_ex-ante_2018, cucurachi_life-cycle_2019} Nevertheless, analysis remains informative because upstream “background” processes (\eg energy supply, solvents, common precursors) are relatively well characterized and often dominate cradle-to-gate impacts of materials.\cite{wender_anticipatory_2014, arvidsson_environmental_2018} An early-stage \ac{LCT} framework should therefore flexibly support multiple \ac{LCIA} methods while allowing for provisional mappings, proxy indicators, or scenario-based characterization where factors are unavailable.

At low \acp{TRL}, assigning precise impact scores or rankings can be misleading, as apparent differences often reflect modeling assumptions rather than real distinctions.\cite{blanco_assessing_2020, prado_challenges_2022} Applying principles of "uncertainty as information" and "decision relevance over precision," the aim shifts from \textit{confirmation} to guiding \textit{exploration} of feasible trajectories within design space.\cite{mendoza_beltran_quantified_2018} Established \ac{LCA} methods---\eg stochastic and sensitivity analyses, scenario comparison, and simple bounding---can be integrated to evaluate outcomes as conditional and distributional rather than fixed.\cite{huijbregts_part_1998, arvidsson_environmental_2018, igosHowTreatUncertainties2019, thonemann_how_2020, heijungs_probability_2024} Emerging probabilistic frameworks further enable representation of low-\ac{TRL} uncertainty, propagation of parameterized inventories, and evaluation across evolving scenarios, supporting identification of hotspots and decision-relevant thresholds with associated confidence.\cite{wender_anticipatory_2014, lacirignola_lca_2017, mendoza_beltran_when_2018, ravikumar_novel_2018, sacchi_prospective_2022, douziech_life_2023, jouannaisENvironmentalSuccessUncertainty2024} Relatedly, early-stage \ac{LCT} is often benchmarked against incumbents, but such comparisons require caution due to \ac{TRL} mismatches and uncertain scale-up;\cite{arvidsson_environmental_2018, woods-robinson_controversy_2025} prospective \ac{LCA} methods help account for learning effects, process improvements, and evolving benchmarks.\cite{caduff_scaling_2014, piccinno_laboratory_2016, weyand_scheme_2023}

When interpreting early-stage \ac{LCT}, it is important to keep in mind both the presence of value-laden choices and the distinction between \textit{informing} versus \textit{making} decisions. \ac{LCA} structures information about potential consequences and tradeoffs, but interpreting results---such as weighting impacts or setting acceptable uncertainty thresholds---necessarily involves contextual, value-laden choices.\cite{iso_iso_2006-1, prado_sensitivity_2020} Early-stage \ac{LCT} can help anticipate and communicate potential implications transparently to inform decision-making, but---as with standard \ac{LCA}---actually \textit{making} decisions remains the responsibility of informed stakeholders.

\subsection{Integration, Interoperability, and Accessibility}

Emerging advances in open data and infrastructure further enable this vision, especially tools following FAIR principles: Findability, Accessibility, Interoperability, and Reuse.\cite{wilkinson_fair_2016} On the engineering side, open databases of materials calculations\cite{curtarolo_aflowliborg_2012, jain_materials_2013, saal_materials_2013, draxl_nomad_2019, horton_accelerated_2025} and chemical data\cite{nist2001webbook, kim_pubchem_2016, kearnes2021ord} provide access to property and process information, while corresponding open software enables analysis.\cite{ong_python_2013, hjorth_larsen_atomic_2017, bento_open_2020, cortes-pena_biosteam_2020, lee_idaes_2021, medeiros_dwsim_2025} Meanwhile, open-source \ac{LCA} frameworks like \texttt{Brightway} support parametric \ac{LCI} construction and dynamic integration with uncertainty analyses tools and other open software,\cite{mutel_brightway_2017, jolivet_lca_algebraic_2021, sacchi_prospective_2022} while open \ac{LCI} databases (the exception, not the norm) provide standardized background systems that can be coupled to chemistry-informed foreground models.\cite{uslci2025lci, bafu2025lci}

Within this ecosystem, \ac{ML} models can function as \textit{translators} across layers--- \textit{not} as a fully-automated replacement to \ac{LCT}. In addition to \ac{ML}-enabled advances across synthesis prediction\cite{karpovich_interpretable_2023, williamson_using_2023, wu_how_2025} and chemical process simulation,\cite{sharma_hybrid_2022, srinivasan_artificial_2025}, data-driven \ac{LCA} methods are rapidly increasing\cite{ghoroghi_advances_2022, romeiko_review_2024, popowicz_digital_2025}, with promising low-\ac{TRL} applications in surrogate process modeling,\cite{martinez-ramon_frameworks_2024} \eg automated \ac{LCI} construction,\cite{kock_automation_2023} scenario generation for prospective \ac{LCA},\cite{blanco_machine_2024} and guiding uncertainty assessment.\cite{akrami_addressing_2025} Data-driven methods could help bridge materials, processes, and inventory modeling within unified workflows, and enable multi-objective exploration of performance, cost, and sustainability impacts under uncertainty.

At the same time, this integration highlights an important boundary: automatically filling data gaps risks optimizing the wrong objective when sustainability priorities are implicit or undefined. Integrating \ac{LCT} into materials discovery therefore requires transparent coupling of models and metrics, with automation supporting interpretation and comparison rather than resolving value-laden tradeoffs. Furthermore, the rapid proliferation of competing \ac{ML} models, limited standardization and unclear validation practices hinder reliable application in decision-relevant \ac{LCA} workflows, and adherence with FAIR principles (\eg cross-domain interoperability) remains an open challenge.\cite{wilkinson_fair_2016} 


\section{A Plan for Coordination Across Fields}

While many building blocks are in place, realizing life cycle–aware materials design at early \acp{TRL} will require coordinated effort across materials science, engineering, assessment, and research infrastructure. We outline below concrete priorities, grouped by stakeholder(s), and emphasize that many of these are not specific to \ac{LCA}, and are complementary with other research priorities.

\begin{itemize}
\item \textbf{Align predictive synthesis with \ac{LCT} relevance}\\
\textit{Stakeholder(s): Computational materials scientists}\\
Extend predictive synthesis tools beyond equilibrium and single-step reactions to capture non-equilibrium pathways, metastability, morphology dependence, purification, and multi-step routes. Expand tools beyond solid-state synthesis routes (\eg sol-gel, solvothermal, microwave, thin film). Standardize synthesis taxonomies and data formats, and benchmark predictive performance so outputs can inform process modeling and early-stage \acp{LCI}.

\item \textbf{Link process modeling to material design data}\\
\textit{Stakeholder(s): Process engineers \& modelers}\\
Build on existing methods to estimate processing energy at low \acp{TRL}, including heating, cooling, mixing, separations, and purification. Better link process models to synthesis routes and material properties, and formalize concept-to-commercial scale-up and learning trajectories. For such assumption-sensitive and cross-disciplinary workflows, open models and data are essential for critical evaluation, interoperability, and cumulative improvement.

\item \textbf{Expand materials-informed inventories}\\
\resizebox{\linewidth}{!}{\textit{Stakeholder(s): LCA data curators \& materials scientists}}\\
Extend \ac{LCI} data structures to represent chemistry-, form-, and especially purity-specific variants of materials, informed by thermodynamic and materials property data. Support parametric inventories that evolve with increasing \ac{TRL}. Advance approaches for handling emerging materials that lack established LCIA characterization factors (\eg proxy mappings, grouping strategies, semi-automated provisional characterization).\cite{cucurachi_ex-ante_2018, holmquist_potential_2018, hou_estimate_2020, von_borries_potential_2023}

\item \textbf{Establish best practices for early-stage LCT}\\
\resizebox{\linewidth}{!}{\textit{Stakeholder(s): LCA/LCT analysts \& methods developers}}\\
Build consensus on protocols for \ac{LCT} of emerging technologies, including uncertainty representation, scale-up modeling, incumbent comparisons, impact category selection, treatment of novel materials without LCIA characterization factors, and decision-informed interpretation appropriate to low \acp{TRL}.


\item \textbf{Build interoperable, open data infrastructures} \\
\textit{Stakeholder(s): Data scientists across fields}\\
Develop open, interoperable data structures and modular workflows linking materials, synthesis, process, and \ac{LCI} databases, supported by standardized taxonomies and shared ontologies that make benchmarking, comparison, and tool integration possible.

\item \textbf{Practice uncertainty-informed, \ac{LCT} at low \acp{TRL}}\\
\textit{Stakeholder(s): Materials designers \& R\&D teams}\\
Integrate provisional \ac{LCT} considerations into early materials selection, synthesis planning, and performance screening. Reframe uncertainty as a design feature: work with bounds, ranges, and feasible design windows to compare options and guide prioritization, even under incomplete knowledge.

\item \textbf{Leverage \& integrate early-stage LCT incentives}\\
\textit{Stakeholder(s): Decision-makers \& funders}\\
Embed \ac{LCT} into funding calls, review criteria, technology due diligence, and project milestones, recognizing competitive advantages \ac{LCT} can bring. Co-integration with \ac{TEA} is particularly tractable, leveraging overlapping process and inventory data needs. Reward uncertainty-aware analysis, transparency, and independent review to reduce bias. Support shared infrastructure and standardized reporting to enable integration across materials, process, and sustainability communities.

\item \textbf{Coordinate targeted cross-disciplinary efforts}\\
\resizebox{\linewidth}{!}{\textit{Stakeholder(s): Funders, institutions \& leaders across fields}}\\
Integration will not emerge through \textit{ad hoc} collaboration alone;\cite{bromham_interdisciplinary_2016} it requires committed champions, explicit incentives, sustained funding, and coordination hubs.\cite{borner_multi-level_2010, cravens_science_2022} Priorities include targeted workshops and facilitated dialogue to build shared language, mentoring and co-advised trainee models to sustain integration, paired funding calls to incentivize joint projects, and stewardship of shared datasets and open models---and learning from past cross-disciplinary successes.\cite{collins_human_2003, wilkinson_fair_2016, eyring_overview_2016, fichtner_rechargeable_2022, horton_accelerated_2025}

\end{itemize}

\section{Conclusion}

Sustainability leverage is greatest at the earliest stages of materials design, when choices about composition and process remain flexible and consequential. This Perspective proposes an epistemic, bottom-up framework for integrating \ac{LCT} into early-stage materials discovery, enabling results to inform design decisions before impacts are locked in. Uncertainty at low \acp{TRL} is not a barrier to action, but rather a source of insight to bound outcomes, identify dominant drivers, and guide experiments. Early-stage \ac{LCT} places agency above precision---using incomplete but structured information to shape materials innovation before trajectories become fixed.

Rather than prescribing specific tools or automated workflows, we intend this framework as a conceptual roadmap to connect fragmented communities and emerging workflows---including synthesis prediction, process modeling, and parametric inventory construction---to enable \ac{LCT} from the point of materials selection and synthesis design onward. Complementary early-stage integrative approaches in green chemistry,\cite{zhu_application_2020, reyes_green_2023, chen_data-driven_2026} energy technologies (\eg batteries\cite{putsche_framework_2023}), and environmental justice (\eg JUST-R\cite{dutta_just-r_2023, arkhurst_evaluating_2023, dutta_applying_2024}) point to broader efforts toward ethical science guided by the precautionary principle.\cite{kriebel_precautionary_2001}

Although our focus has been primarily environmental impacts through \ac{LCT} and \ac{LCA}, the modular structure of the framework could support additional forms of evaluation, including social \ac{LCA}, \ac{TEA}, or sustainable-by-design approaches. More broadly, the principles outlined here should extend across materials systems and research workflows. Realizing this vision will require closer coordination across materials science, sustainability analysis, process engineering, and decision science to align methods, share data, and develop interoperable tools---so that \ac{LCT} informs materials innovation from the outset rather than through downstream correction.

\section{Acknowledgments}

The authors thank Celeste Melamed, Evan Spotte-Smith, Lizzie Mahoney, Michael Machala, and Taylor Uekert for insightful discussions and constructive feedback during the preparation of this
manuscript. R.W.R. was supported by a Distinguished Postdoctoral
Fellowship at the University of Washington’s Clean Energy Institute. This work was supported by funding from Toyota Research Institute.

\printbibliography

\end{document}